\documentclass[article,onefignum,onetabnum]{siamart171218}
\usepackage{sidecap}
\usepackage{pdflscape}
 \DeclareMathOperator*{\argmax}{argmax}

\usepackage{lipsum}
\usepackage{amsfonts}
\usepackage{graphicx}
\usepackage{epstopdf}
\usepackage{algorithmic}
\ifpdf
  \DeclareGraphicsExtensions{.eps,.pdf,.png,.jpg}
\else
  \DeclareGraphicsExtensions{.eps}
\fi

\newsiamremark{remark}{Remark}
\newsiamremark{hypothesis}{Hypothesis}
\crefname{hypothesis}{Hypothesis}{Hypotheses}
\newsiamthm{claim}{Claim}

\headers{Black Litterman and ESG Portfolio Optimization}{A. Alpern, S.T. Rachev}

\title{Black Litterman and ESG Portfolio Optimization\thanks{Submitted to the editors DATE.
}}

\author{Aviv Alpern\thanks{Bravo Risk Management
  (\email{avivalpern@gmail.com}).}
\and Svetlozar Rachev\thanks{Department of Applied Mathematics, Texas Tech University, Lubbock, TX 
  (\email{zari.rachev@ttu.edu}).}
}

\usepackage{amsopn}

\makeatletter
\newcommand*{\addFileDependency}[1]{
  \typeout{(#1)}
  \@addtofilelist{#1}
  \IfFileExists{#1}{}{\typeout{No file #1.}}
}
\makeatother

\newcommand*{\myexternaldocument}[1]{%
    \externaldocument{#1}%
    \addFileDependency{#1.tex}%
    \addFileDependency{#1.aux}%
}

\ifpdf
\hypersetup{
  pdftitle={Black Litterman and ESG Portfolio Optimization},
  pdfauthor={A. Alpern, S. Rachev}
}
\fi

\myexternaldocument{supplement}

\begin{document}

\maketitle

\begin{abstract}
We introduce a simple portfolio optimization strategy using ESG data with the Black-Litterman allocation framework. ESG scores are used as a bias for Stein shrinkage estimation of equilibrium risk premiums used in assigning Black-Litterman asset weights. Assets are modeled as multivariate affine normal-inverse Gaussian variables using CVaR as a risk measure. This strategy, though very simple, when employed with a soft turnover constraint is exceptionally successful. Portfolios are reallocated daily over a 4.7 year period, each with a different set of hyperparameters used for optimization. The most successful strategies have returns of approximately 40-45\% annually. 
\end{abstract}


\section{Introduction}
Socially responsible investing has become a common topic for research, specifically with interest in its application to asset allocation. Recently, many studies have worked to elucidate the impact of the components of ESG ratings on assets and firms. It's suggested that positive indicators of these components are linked to reduced risk during periods of potential drawdowns \cite{FM19, NV14, Paul17}. While there is no shortage of literature on the implications of ESG factors on risk and market behavior, we do not present this topic in depth and instead point the reader to Ielasi et al. \cite{ICZ20} for a more comprehensive view. Here, we instead investigate a method for incorporating ESG scores in smart beta portfolio allocation.

A common method of asset allocation familiar to portfolio managers is the Black-Litterman (BL) model for incorporating investor's market views \cite{BL91}. This allows adjusting expected return of individual assets based on a portfolio manager's belief of asset returns or relative performance between assets. We blend the BL model with ESG data, allowing both investor views and the ESG factor approach to directly affect the estimated mean of the distribution of portfolio returns. Under the original BL assumptions, the market is modeled using normal distributions which is known to underestimate the probability of unlikely events \cite{Kim11, Mandelbrot63}. For incorporating the BL model views in calculating tail risk measures of elliptically contoured distributions, see Rosella et al. (2007) \cite{stableBL}. Here, we only use BL to model expected returns, while using the normal-inverse Gaussian (NIG) to approximate tail-risk in the form of conditional value-at-risk (CVaR).

In our empirical analysis, we compare the performance of portfolios comprised of different levels of both risk aversion and ESG weighting. Each portfolio is analyzed over a 1175 day testing period (2017-2021), with 4 years of data used for model fitting. The performance is then compared using different reward risk ratios to quantify results.

\section{Modern Portfolio Theory}
\label{sec:theory}

\subsection*{Black-Litterman}\footnote{For a in-depth explanation see Meucci (2010) \cite{Meucci10}, and , Rachev ST, Hsu JSJ, Bagasheva BS, Fabozzi FJ.\textit{Bayesian Methods in Finance} \cite{BMF}} 
For a universe of M assets, we begin by assuming that the observations of asset returns $X$ follow a normal distribution:
\begin{equation}\label{eq:2.1}
X \sim \boldmath{N}(\mu,\Sigma)
\end{equation}
where $\mu \in \mathcal{R}^M$ and $\Sigma \in \mathcal{R}^{MxM}$ are the multivariate mean and covariance respectively. It is common to use the sample covariance as an estimate for $\Sigma$ but we use another approach described in a latter section. 

The parameters $\boldmath{\mu}$ and $\boldmath{\Sigma}$ are themselves random variables. Thus, we don't know their exact values. As suggested by Satchell and Scowcroft \cite{SS2000} it is standard to apply the Bayesian approach of predictive inference and use the informative prior:
\begin{equation} \label{eq:2.2}
\mu \sim \boldmath{N}(\Pi,\tau\Sigma).
\end{equation}

\noindent Here, the extra parameters $\Pi$ and $\tau$ come from the \textit{Capital Asset Pricing Model} (CAPM). $\tau$ represents the uncertainty of the accuracy of the CAPM assumption and $\Pi$ is the equilibrium risk premium, defined as:
\begin{equation}\label{eq:2.3}
\boldmath{\Pi = \delta\Sigma\omega_{eq}}.
\end{equation}
$\delta$ is a risk aversion parameter, which is the Sharpe ratio divided by the standard deviation, i.e. ${(R_M - R_f)}/{\sigma^2}$, the expected asset return and risk free return being denoted by $R_M$ and $R_f$ respectively. As the name of the model suggests, the equilibrium portfolio weights $\omega_{eq}$ are the relative market capitalization of the portfolio components. 

\subsection*{Market Views} Within this Bayesian framework, it follows that we should allow portfolio managers to input a priori beliefs into the model. These beliefs categorized as \textbf{relative views}: in which we say asset A is expected to outperform asset B by 5\% or \textbf{absolute views}: in which we specify the expected return of asset C at the end of the year. For a full explanation and example of inserting market views see Meucci (2010) \cite{Meucci10}. Our views are expressed through transformation of our original random variables $\mu$ and $\Sigma$ giving the posterior mean, $\mathbf{\mu_{BL}}$, and covariance matrix, $\mathbf{\Sigma_{BL}^\mu}$

\begin{equation}\label{eq:2.4}
    \boldmath{\mu_{BL} = ((\tau \Sigma)^{-1} + P^T\Omega P)^{-1} \hspace{1mm} ((\tau \Sigma)^{-1}\pi + P^T \Omega v)}
\end{equation} 

\begin{equation}\label{eq:2.5}
    \boldmath{\Sigma_{BL}^{\mu}  = ((\tau \Sigma)^{-1} + P^T \Omega P)^{-1}}
\end{equation}

If we are to write, P = (1, 0; -1, 1), $v = (0.05, 0)^T$, and $\Omega$ = diag(.0001, .01), then we are expressing that we believe (1) the the return of asset A to be 5\% at the end of the year with a relatively high confidence and (2) that asset A and B are expected to have the same return with low confidence. As in any statistical model, the added hyperparameters ($\tau$, P, $\Omega$, v), are far from trivial to designate which attests to the skill of the portfolio manager to do so. 

\subsection*{Risk Measure: CVaR} Under mean-variance portfolio optimization, we can find a closed-form solution to the portfolio providing the highest Sharpe ratio. In this case we use the above model to pick the portfolio weights that maximize the posterior mean subject to minimizing the risk implied by posterior variance. However, since we instead use CVaR as our risk metric, we use the method outlined by Rockafellar and Uryasev \cite{cvar} defined below.

We start by defining value-at-risk (VaR), i.e. the return that defines the $\beta$ percentile of the loss function:
\begin{equation}\label{eq:2.6}
    VaR_{\beta} = \min\{A \in \mathcal{R}: \Psi(\omega,A) \geq \beta\}
\end{equation}
\begin{equation}\label{eq:2.7}
    \Psi(\omega,A) = \int_{f(\omega,x) \geq A}p(x)\hspace{1mm}dx 
\end{equation}
\noindent Where the loss function, $f(\omega,x)$, is the loss of a portfolio composed of asset vector x and weight vector $\omega$. The function $\Phi(\omega,A)$ is the probability of the loss not exceeding threshold return of A. For a fixed portfolio and thus constant $\omega$, $\Phi(\omega,x)$ is the cumulative distribution of the portfolio's returns.

The VaR metric has itself been used to quantify portfolio risk. However, it is not a coherent risk measure \cite{risk_book} so it's use is being curtailed in relevant literature. The distinctive property which VaR fails to meet is the subadditivity of risk between two or more assets. It's possible for the VaR of two assets to be greater than that of the sum of each asset. What we would like is to expect that the hedging the investment reduces risk as opposed to increasing it. For more information on requirement of coherent risk measures see Artzner et al. \cite{coherent_risk}. The need for a coherent risk metric led to the development of CVaR which we can now define to be:
\begin{equation}\label{eq:2.8}
    CVaR_\beta = \frac{1}{1-\beta} \int_{f(\omega,x) \geq VaR_\beta} f(\omega,x)\hspace{1mm}p(x)\hspace{1mm}dx
\end{equation}
\noindent which is the expected value of the loss of a portfolio when the loss exceeds $VaR_\beta$.

The evaluation of the above equation can make its optimization nontrivial. However, Rockafellar \& Uryasev show that instead we can optimize the function,
\begin{equation}\label{eq:2.9}
F_\beta(\omega,VaR) = VaR + \frac{1}{1-\beta}\int_{x \in \mathcal{R}^N}[f(\omega,x) - VaR]^+ \hspace{1mm} p(x) \hspace{1mm} dx
\end{equation}
\noindent with $[\hspace{1mm}]^+$ denoting $max(0,g)$ for some function g. The above equation retains the convexity property we need and simplifies the optimization process and returns the global minimum. This result comes from the fact that
\begin{equation}\label{eq:2.10}
    CVaR_\beta = \min F_\beta(\omega,VaR).
\end{equation}
\noindent The minimization process just outlined leads to the definition of portfolio optimization process shown in \cite{portmax}:
\begin{equation}\label{eq:2.11}
    \omega =\underset{\omega}\argmax{\hspace{1mm}\{\alpha\hspace{0.7mm} \omega^T \hspace{0.7mm} \mathbf{\mu_{BL}} - (1-\alpha) \hspace{1mm} CVaR\}}
\end{equation}
\noindent The risk aversion parameter, $\alpha \in [0,1]$, is used to adjust the weighting of CVaR relative the the expected return. This way, the ability of the investor to assume risk is included in the portfolio choice. 

\subsection*{Risk Modeling: Heavy Tailed Distributions}

The loss function,\\ $f(\omega,x)$ in Eq (\ref{eq:2.7} - \ref{eq:2.9}) is dependent on the assumed distribution of asset returns. As discussed earlier, the normal distribution, assumed in the Black-Litterman model, is unable to accurately represent tail risk. To model tail-risk we use the normal-inverse Gaussian (NIG) distribution which has semi-heavy tails. Introduced by Barndorff-Nielsen in 1977 \cite{GH1}, the probability density function of NIG is:

\begin{equation}\label{eq:2.12}
    f(x) = \frac{\alpha\delta \mathbf{K_1}(\alpha\sqrt{\delta^2+(x-\mu)^2})}{\pi \sqrt{\delta^2 + (x - \mu)^2}}e^{[\delta\sqrt{\alpha^2 + \beta^2} + \beta(x-\mu)]} 
\end{equation}

\noindent with parameters denoting \vspace{1mm}

\begin{itemize}
    \item[] $\alpha$ = tail-heaviness 
    \item[] $\beta$ = asymmetry
    \item[] $\delta$ = scale
    \item[] $\mu$ = location
\end{itemize}

\vspace{1mm}

\noindent and $\mathbf{K_1}$ representing the modified Bessel-function of the third kind. While the original paper gives the superclass of NIG, the class of generalized hyperbolic distributions, the special case of NIG is closed under convolution \cite{GH2} making it useful for derivative pricing \cite{magh3}.

For the ease of fitting the distribution, we instead use a modification on the generalized hyperbolic distribution, namely the multivariate affine generalized hyperbolic (MAGH) distribution \cite{MAGH1} with the process outlined by Fajardo \& Farias \cite{MAGH2}. Although, we specifically use the subclass multivariate affine NIG, we will continue to refer to this as MAGH. A multivariate random variable $\mathbf{X} \in \mathcal{R}^M$ is a MAGH random variable if \vspace{-7mm}

\begin{equation}\label{eq:2.13}
    \mathbf{X \stackrel d= AY + m}
\end{equation}

\noindent where $\mathbf{Y} \in \mathcal{R}^M$ is a vector of univariate generalized hyperbolic random variables, $\mathbf{A} \in \mathcal{R}^{M \times M}$ is a decomposition of a the covariance matrix $\mathbf{\Sigma}$ s.t. $\mathbf{AA^T = \Sigma}$ and $\mathbf{m} \in \mathcal{R}^M$. Note that $\mathbf{Y}$ itself is not a generalized hyperbolic distributed variable. 

Now we have the required definitions and notation to describe our method of computing the optimal portfolio.

\section{Methods}
\label{sec:methods}
For a matrix of N observation of M assets, $\mathbf{X}\in \mathcal{R}^{N\times M}$ we weight the returns such that the i-th column of $\mathbf{X}$ is weighted as so: \vspace{-5mm}

\begin{equation}\label{eq:3.1}
    \mathbf{W_i} = (1-\lambda)\mathbf{X_i} + \lambda \xi_i
\end{equation}

\noindent with the ESG score of the i-th asset denoted as $\xi_i$. This, in effect, gives the Stein shrinkage of the mean using ESG scores as the bias:
\begin{equation}\label{eq:3.2}
    m_i = (1-\lambda)\mu_i + \lambda\xi_i
\end{equation}
where m is the expected mean of returns.

The transformation of the correlated data $\mathbf{X}$ onto the set of uncorrelated variables is achieved via the Cholesky decomposition of the sample covariance matrix. Writing the Cholesky decomposition of $\mathbf{\Sigma}$ as $\mathbf{R}$ \textit{s.t.} $\mathbf{R^TR = \Sigma}$, we get the relation
\begin{equation}\label{eq:3.3}
    \mathbf{B} = Chol(Cov(\mathbf{W})) = (1-\lambda)\mathbf{R}
    \vspace{-2mm}
\end{equation}

\noindent leading to
\vspace{-.5mm}
\begin{equation}\label{eq:3.4}
\begin{split}
    \mathbf{y} &= \mathbf{(wB^{-1}-m)D^{-1}} \\
    \mathbf{y} &= (\mathbf{wR^{-1}}\frac{1}{1-\lambda} - [(1-\lambda) \boldsymbol{\mu} + \lambda \hspace{0.5mm} \boldsymbol{\xi}])\mathbf{D^{-1}}
\end{split}
\end{equation}

\noindent The matrix $\mathbf{D}$ is a diagonal matrix of the scale variables, $\delta$, of the marginal NIG distributions. Likewise, $\boldsymbol{\mu}$ is a row vector containing the location parameters of the marginals. The lower case representation, $\mathbf{w} \in \mathcal{R}^{1\times M}$, is to denote the M-dimensional random variable as opposed to the matrix of observations. The new variable introduced here, $\mathbf{y} \in \mathcal{R}^{1\times M}$, is a random vector where every variable has mean \{$\vec{0}$\} and diagonalized covariance matrix.

To calculate the marginal parameters $\mu$ and $\delta$ at time $t$ we use the autoregression model ARMA(1,1)-GARCH(1,1) \cite{GARCH}: \vspace{-5mm}

\begin{subequations}
\begin{equation}\label{eq:3.5a}
  \mathbf{X}_{t} = p + \phi \hspace{.5mm}\mathbf{X}_{t-1} + \theta \hspace{.5mm} \epsilon_{t-1} + \epsilon_t
\end{equation}    
\vspace{-5mm}
\begin{equation}\label{eq:3.5b}
  \sigma^2_{t} = q + \phi \hspace{.5mm}\epsilon^2_{t-1} + \gamma \hspace{.5mm} \sigma^2_{t-1} 
\end{equation}
\end{subequations}
\noindent where the $\epsilon$ is a error term at time $t$, and $\mu_t = \mathbf{X_t} - \epsilon_t$. The parameters of the model are fit using MLE under the assumption that the innovation series, $\epsilon_t$, has a standard normal distribution. These models have shown to be successful proxies for calculating portfolio risk when used with a multitude of distributional assumptions on the innovations \cite{kim16,kim2020portfolio,Kim11}. The consequence of using the autoregressive model, beyond capturing autocorrelation and volatility clustering in the data, is that we need only fit the NIG parameters $\alpha$ and $\beta$ on the innovations. 

This process is used to calculated the portfolio CVaR, where the weighted probability densities supplies our loss function introduced in Eq (\ref{eq:2.7}). It should be noted that NIG is only closed under convolution if the distributions being convolved have the same $\alpha$ and $\beta$, which is not the case. The 'mixing' of the random vector, $\mathbf{y}$, by matrix $\mathbf{B}$ in Eq (\ref{eq:3.4}) leads to this the convolution. Therefore, there is no closed form solution to the distribution of $\mathbf{w}$ or the predicted distribution of returns $\omega^T \mathbf{w}$. Instead, we can approximate the distribution through properly scaling and mixing samples taken from, $\mathbf{y}$. With this approximation, the objective function in Eq (\ref{eq:2.9}) for a sample of $q$ observations becomes
\begin{equation}\label{eq:3.6}
    F_{\beta}(\omega,VaR_\beta) \approx VaR_{\beta} + \frac{1}{q(1-\beta)}
    \sum_{j = 1}^{q}[\omega^T \mathbf{w}_j - Var_\beta]^+
\end{equation}
where $\mathbf{w}_j$ is the transformation of the j-th observation of $\mathbf{y}$ as defined in Eq (\ref{eq:3.4}) \cite{cvar}.

We consider portfolios allocated from two different approaches: the standard mean-CVaR optimization and the Black Litterman $\boldsymbol{\mu_{BL}}$-CVaR optimization. Instead of using the market capitalization from CAPM to weight our assets, we instead use the current Dow Jones composition weights of each asset. In this way the method for determining $\boldsymbol{\omega}_{eq}$ is similar to Eq (\ref{eq:3.1}):
\begin{equation}\label{eq:3.7}
    \omega_{\xi,i} = \frac{\xi_i}{\sum_{j=1}^{M}\xi_j} 
    \hspace{9mm}
    \omega_{C,i} = \frac{C_i}{\sum_{j=1}^{M}C_j}
\end{equation}
\begin{equation}\label{eq:3.8}
    \omega_{eq,i} = (1-\lambda)\omega_{C,i} + \lambda_{\xi,i}
\end{equation}
letting $C_i$ and $\omega_{eq,i}$ be the Dow Jones weight and equilibrium weight for the i-th asset respectively.

At time t, we choose our portfolios weights to maximize:
\begin{equation}\label{eq:3.9}
    \omega_t = \underset{\omega}\argmax{ \{ \hspace{1mm} \alpha \hspace{0.7mm} \omega^T R_t -  (1-\alpha)CVaR_t  - \rho \hspace{.7mm} \mathbf{||}\omega_{t} - \omega_{t-1} \mathbf{||}_1  \}}
\end{equation}
Here, $R_t$ is the predicted return of our assets at time t, and CVaR is calculated at the given level $\beta$ and induced by the weight vector $\omega$. The last term on the right hand side is a soft turnover constraint, $\rho$, denoting the turnover penalty and $||\cdot||_1$ being the $L_1$ vector norm. 

The hyperparameters from Eq (\ref{eq:3.8} \& \ref{eq:3.9}) that we test for our portfolios construction are $\lambda = [0, 0.25, 0.5, 0.7]$, $\alpha = 0:0.1:1$, and $\rho = [5, 10, 15, 20, 30, 40]\times10^{-4}$ with CVaR calculated at 95\% and 99\%. 

\section{Results}
    In total, we present results for 616 portfolio strategies. As the related figures and performance metrics are somewhat substantial, we introduce results and the relating discussion together to avoid an otherwise unclear presentation of disparate information. We leave the more general discussion of the findings to the later section.
\subsection{Data}
For our analysis we use daily Bloomberg data of the Dow Jones 30 between Jan 3, 2013 and Sep 3, 2021. However, of those 30, we only had data for 29. The asset missing from the data set was Dow Inc (DOW). The ESG data is yearly quotes from Robeco, released on the last trading day in December each year over the period from 2016 to 2020. The data set lacks ESG scores for Walt Disney (DIS) before 2019. Consequently DIS is not included in the portfolio for years without ESG data. Our benchmark is thusly an adjusted DJ index using it's current asset weights, re-normalized so they add to 1 on the remaining 28 asset for the entire period. It should be noted that removing DIS and DOW from the benchmark severely increased it's performance over the period, seeing that these two companies have been amongst the lowest earners of the DJ 30 in the last 5 years. For fitting our model we use a window size of 1007 returns over a period of 1175 days with our first asset return prediction on Jan 3, 2017. The initial portfolio weights for each strategy is composed of the weights defined in Eq (\ref{eq:3.8}).

To assess each portfolio strategy we use reward-risk ratios (RRR) listed in Cheridito \& Kromer (2013) \cite{RRR}. We utilize the Sortino, Gini, and STARR ratios which the display the four properties of monotonicity (M), quasi-concavity (Q), scale invariance (S), and distribution dependency (D) presented in the original paper. All reward-risk ratios are calculated using empirical distributions over the same window size used in training the model, stated above. While this can be sufficient to capture the behavior of the strategy of the period as a whole, to observe sensitivity to short-term market fluctuations would require using the parametric distribution for the predicted daily return. This, however, will be the subject of another paper allowing a more comprehensive evaluation of the performance of the model during different market periods.

\subsection{Computational Results}
The focus of this investigation is to introduce and assess a basic framework for using the Black-Litterman (BL) approach on an asset class with ESG data. Of course, the BL method is tailored for incorporating a portfolio managers views on the market, views that would be proprietary if even moderately successfully. We do not attempt to express market views and instead only present the basic structure of BL on top of CAPM. In this way, we have a simple bias for shrinkage estimation of mean returns. Even without imposing market views, the BL model outperforms the benchmark.

The performance metrics for each portfolios strategy with CVaR95 are shown in Table 2A. The standard mean-CVaR approach shown in Fig.\ref{fig:1} and Table \ref{C95_std05} was unsuccessful compared to the benchmark, having substantially lower returns and RRR. Strategies that had large ESG weights  and $\alpha$, specifically \{$\lambda = 0.25, \hspace{0.8mm} \alpha \geq 0.9$\}, \{$\lambda = 0.5, \hspace{0.8mm} \alpha \geq 0.8$\}, and \{$\lambda = 0.75, \hspace{0.8mm} \alpha \geq 0.5$\}, do not rebalance their portfolios at any time step. These strategies have the best performance out of all standard mean-CVaR approaches and slightly under perform the benchmark in both return and RRR. The relative success of these inert portfolios displays the utility of both the CAPM and ESG together.

For mean-CVaR portfolios with nonzero turnover, the degree of yearly turnover is associated with lower risk and lower reward. Portfolios with \{$\lambda = 0, \alpha \leq 0.5$\} and \{$\lambda = 0.25, \alpha \leq 0.4$\} have the lowest drawdowns of all employed strategies. However, the comparatively meager annual returns of 15-20\% result in low RRR. Among these strategies, increased turnover is directly associated with lower returns. This is typical, as CVaR is sensitive to the data used in fitting. Moreover, the convergence of the distribution is dependent on the tails, converging asymptotically slower for heavier tails \cite{CVaR_sensitivity}.

In contrast to the acceptable level of turnover observed for standard mean-CVaR portfolios with $\rho = 5 \times 10^{-4}$, a majority of CVaR95 BL portfolios with the same constraint had excessive turnovers that made the strategies infeasible. The high turnover is a consequence of the overestimation of returns by the BL model leading to over-sensitivity of the optimal portfolio found in Eq (\ref{eq:3.9}) to the the conditional mean from ARMA \ref{eq:3.5a}. Similar to the results in the mean-CVaR case, small $\alpha$ and large $\lambda$ curtailed the excess turnover. However, the portfolios that had acceptable annual turnover $\leq 3$ still fail to approach the benchmark, the best of which having an annual return of $\approx 27\%$. Generally this would be more than acceptable returns, although the benchmark returned over 31\% annually and had higher RRR. The success of the benchmark can be attributed, at least in part, to the accommodating fed policy following the COVID related market crash of march 2020.

The sensitivity of the BL model could be remedied by adjusting the risk-aversion parameter from Eq (\ref{eq:2.3}), however, we simply increase the turnover constraint. Of the BL strategies that had acceptable turnover, the largest portfolios returns were between 40-45\% annually. These portfolios had high RRRs equal to or greater than the benchmark. These strategies worth noting are labeled and plotted against the benchmark in Fig. \ref{fig:1}.

The results for CVaR99 are listed in Tables 3A-3G. The performance compared with CVar95 was subpar. Strategies that had acceptable turnover rarely approach the benchmark. Many of the strategies that do, analogous to the CVaR95 case, have zero turnover. All zero-turnover strategies have performance metrics slightly below the benchmark. The highest performing portfolios are plotted in Fig. \ref{fig:1}. Across all CVaR99 portfolios, the RRR and drawdown ratio is not significantly higher than that of CVaR95. Points on the tail of a distribution become more sensitive to the observed window of data the distance from the mean increases \cite{CVaR_sensitivity}. Thus it can be hard to hedge against daily fat tailed risk for large percentiles. Consequently, the difficulty of the CVaR minimization problem can be exacerbated by illiquidity. Much of the utility of CVaR as a risk metric is to estimate the amount of cash required in a margin account to prevent a margin call.  

\section{Discussion}
Our Black-Litterman ESG model performs exceptionally well compared to our benchmark, or any well-known composite indexes. Of course, we tested the model with many combinations of hyperparameters. It would be appropriate to cross validate the performance with an out-of-sample data set. However the ESG data for Robeco only goes back to the Dec 30, 2016, so there is not sufficient data to analyze the performance over. Additionally, the characteristics of the market over past 3 years have been very unique. Observing model behavior over only a small subset would be largely subject to the characteristics of the period.  

One of the aforementioned properties of ESG strategies is the reduction of systematic risk during bear markets. We do not directly observe this. During the period of data used, we observe the market before, during, and after the 2020 COVID-19 market crash. As the crash is not an extended period of economic recession, the empirical reward-risk ratios calculated over 4 years is not sensitive enough express the behavior over the rather brief duration. Thus we do not observe any significant under or overperformance. This would be specific to the standard mean-CVaR as the performance of the BL ESG model would be highly sensitive to the views expressed, and thus the views should reflect the market at the given time. In our model, ESG ratings do not directly affect the tail risk of the distribution of predicted returns \footnote{For more information on using views to get get a posterior estimate for CVaR see Giacometti et al. (2007) \cite{stableBL}}. Moreover, both the CAPM and ESG weights included in BL model are expressed on a yearly basis. Accordingly, during the relatively longer bull market period before and after the 2020 crash the BL ESG model clearly performs well above the benchmark as seen in Fig. \ref{fig:1}.

It would be of interest to examine the affects of different political events of the past 5 years on ESG based portfolios. Specifically the time period surrounding the UN climate change conferences and the change of presidential administration in the US. The Trump administration, in particular, had a large affect on the environmental policy in the country. One might expect that the performance of high ESG assets may change as a result of related legislation. This may be of greater scrutiny in later studies.

Additionally, the ESG data used here is yearly. Monthly data exists but is not in our data set. In both cases, the amount of data is relatively small. Giving an estimate for a prediction or uncertainty of such a time series can be nontrivial. Furthermore, the factors that feed into the ESG score provided by each data source are opaque. It may be of use to have a Black-Litterman type model applied to the expectation of ESG. A portfolio manager could then express their views as a result of external information and the certainty of those views compared to data itself. Thus, the certainty, especially for yearly data, can be changed as the time from data point is increases. Nevertheless, the utility of ESG in portfolio management is significant and current techniques would benefit from further research. 

\bibliographystyle{siamplain.bst}
\bibliography{references}

\vspace{1cm}

\appendix{Appendix}

\renewcommand{\figurename}{Fig.}
\renewcommand{\thefigure}{1}
\begin{figure}[h]
    \begin{center}
    \makebox[\textwidth][c]{\includegraphics[width = 17cm]{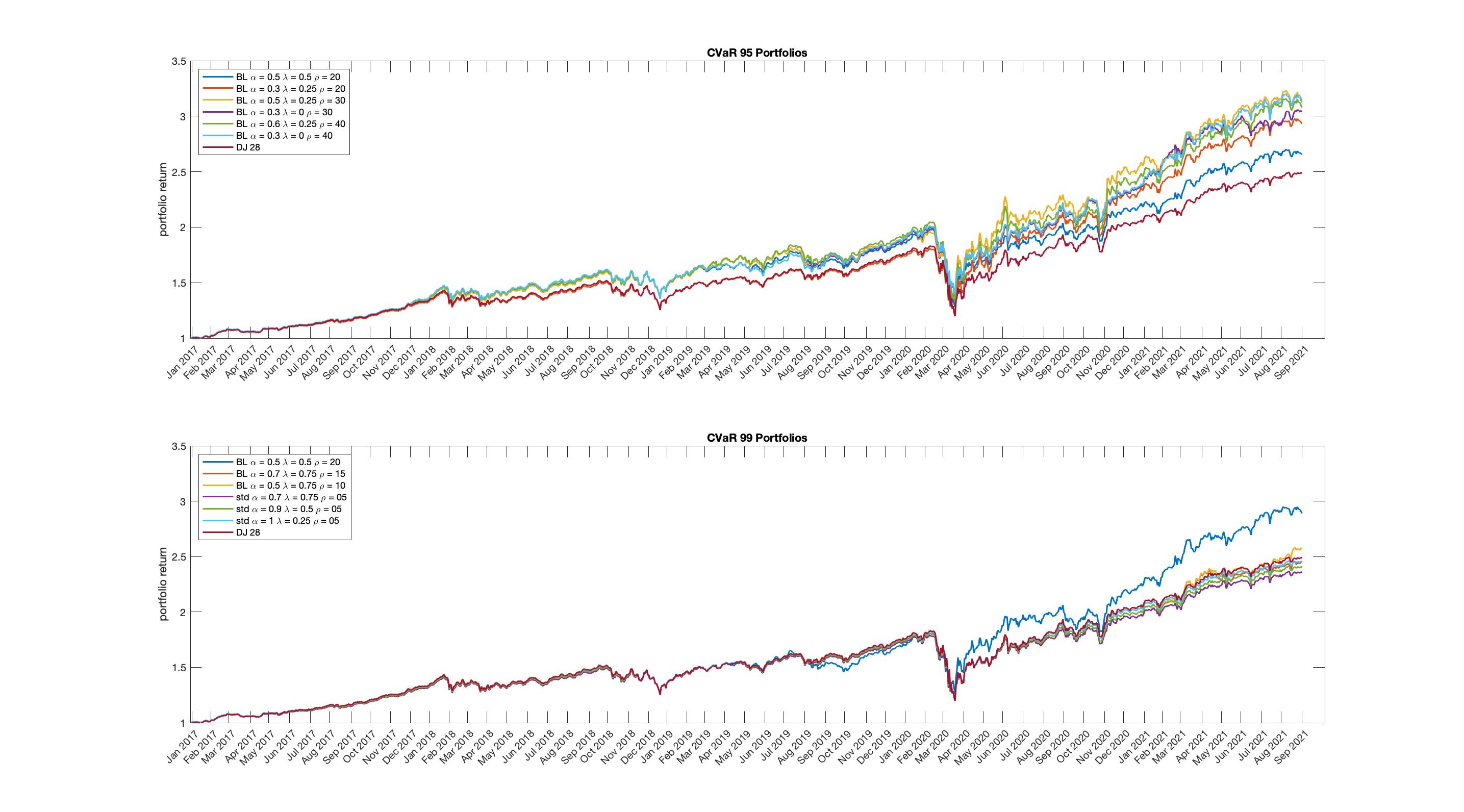}}
    \caption{Comparative plots of the best performing portfolios.}
    \label{fig:1}
    \end{center}
\end{figure}

\vspace{1cm}

\begin{table}[h]
\centering
\begin{tabular}{||c c c c c c c||}
    \hline
    Total Return & Annual Return & Sharpe & Sortino & Gini & STARR & DDR\\
    \hline
    1.4864 & 0.3187 & 0.0743 & 0.1057 & 0.1057 & 0.0288 & 4.538\\
    \hline
\end{tabular}\\
\vspace{4mm}
\caption{Performance metrics of the DJ 28 benchmark. }
\label{}
\end{table}

\renewcommand{\figurename}{Table}
\renewcommand{\thefigure}{2A}
\begin{figure}[b]
    \centering
    \caption{Tables 2A-2G display performance metrics of each portfolio strategy using CVaR95. In Table 1A the row label 'std' denotes the standard mean-CVaR method of portfolio optimization as opposed to the BL method. The Sharpe, Sortino, Gini, and STARR ratios are daily averages, the drawdown ratio (DDR) is calculated using the return and max drawdown over the entire period, and the turnovers are yearly averages. The listed $\rho$ in the row labels is on the order of $1 \times 10^{-4}$. Color maps for each column, except turnovers, are a linear gradient between the min and max of each column across all rows of tables 1A-1G, yellow being the max. For turnovers, the color map is inverted and linear between 0 and 10, 0 being yellow and entries $\geq$ 10 being purple. The color map for all plots is shown beside figure 2A.}
    \includegraphics[width = 19.5cm, angle = 90]{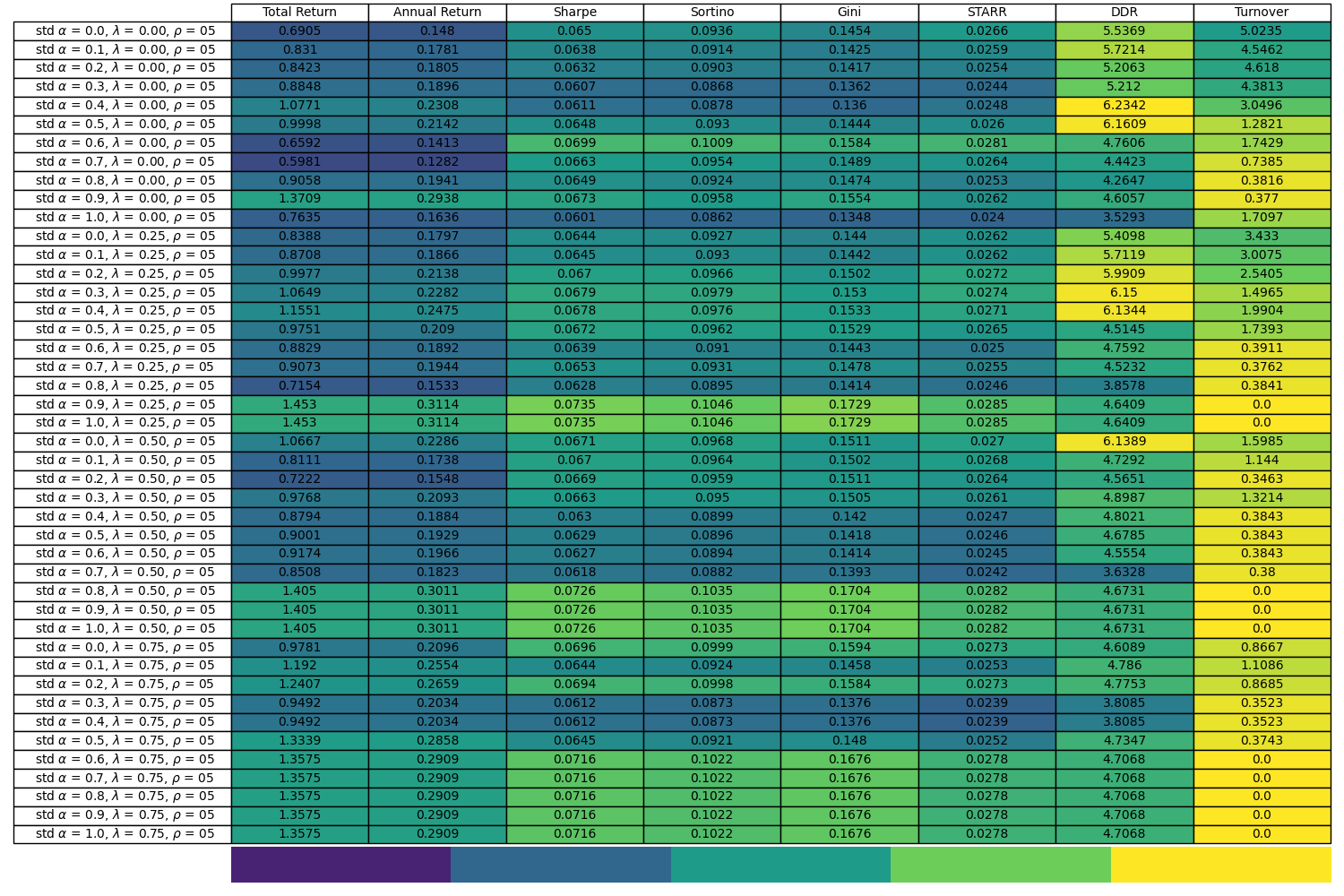}
    
    \label{C95_std05}
\end{figure}

\renewcommand{\thefigure}{2b}
\begin{figure}[b]
    \centering
    \includegraphics[width = 19.5cm, angle = 90]{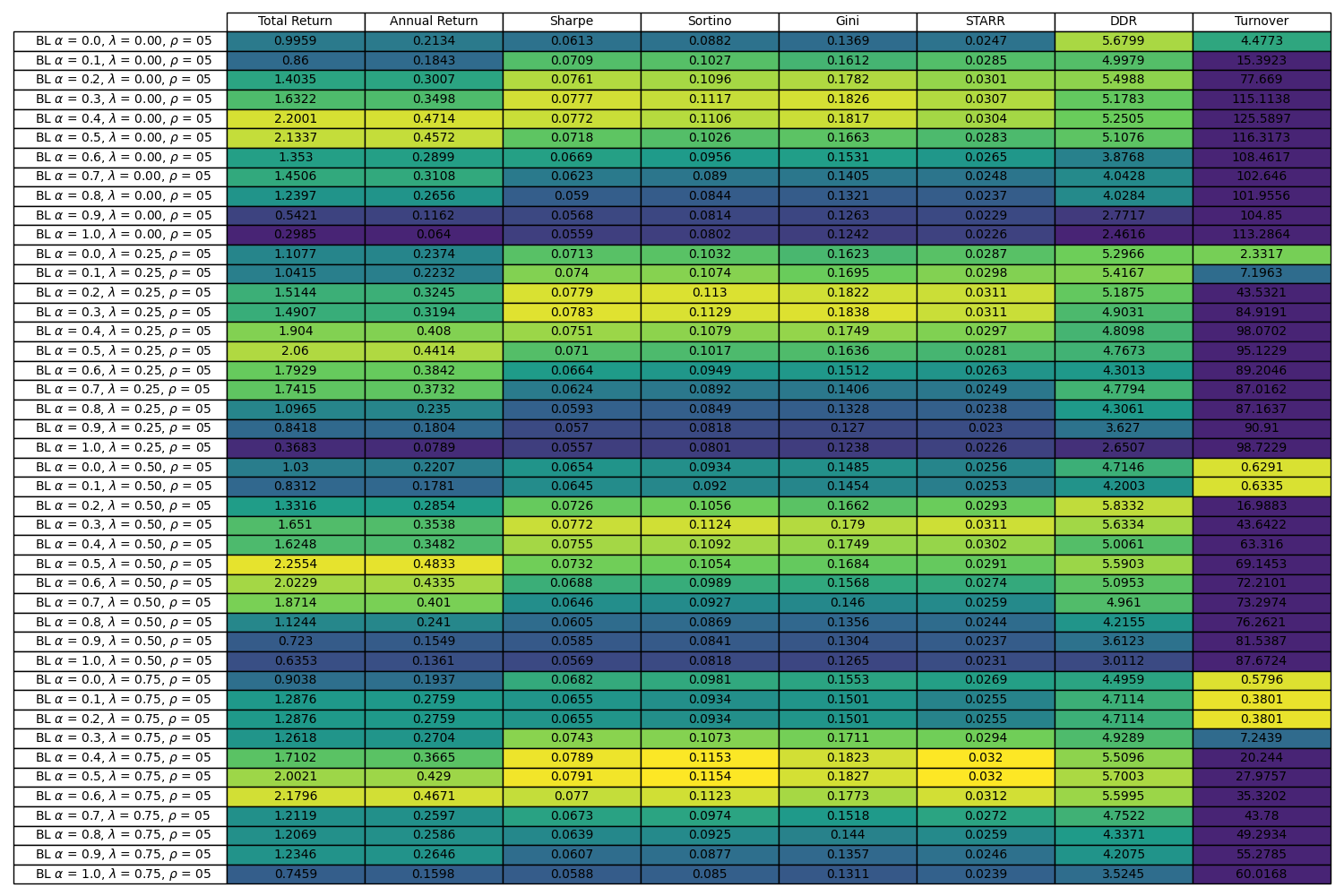}
    \caption{}
    \label{C95_BL05}
\end{figure}

\renewcommand{\thefigure}{2c}
\begin{figure}[b]
    \centering
    \includegraphics[width = 19.5cm, angle = 90]{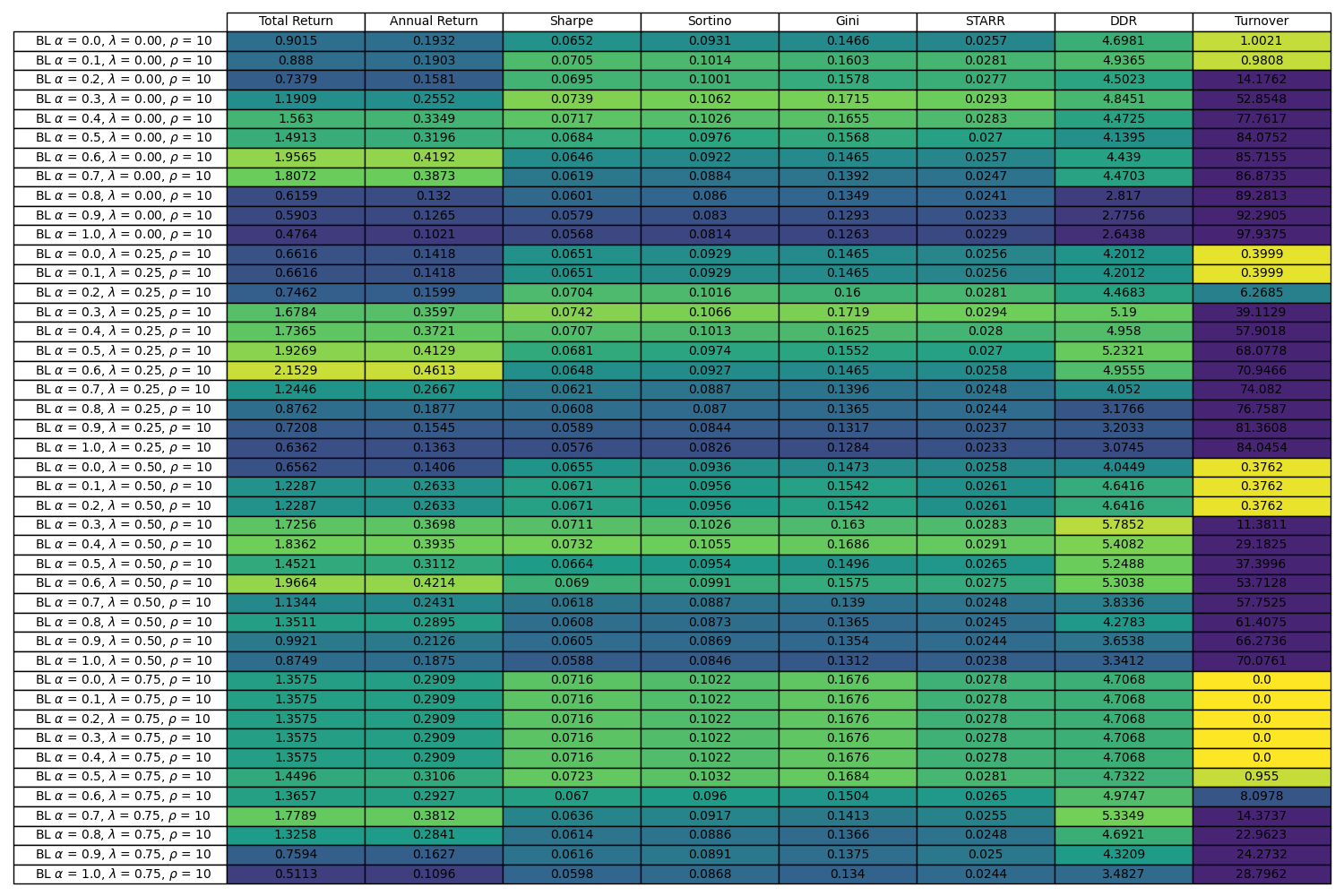}
    \caption{}
    \label{C95_BL10}
\end{figure}
    
\renewcommand{\thefigure}{2d}
\begin{figure}[b]
    \centering
    \includegraphics[width = 19.5cm, angle = 90]{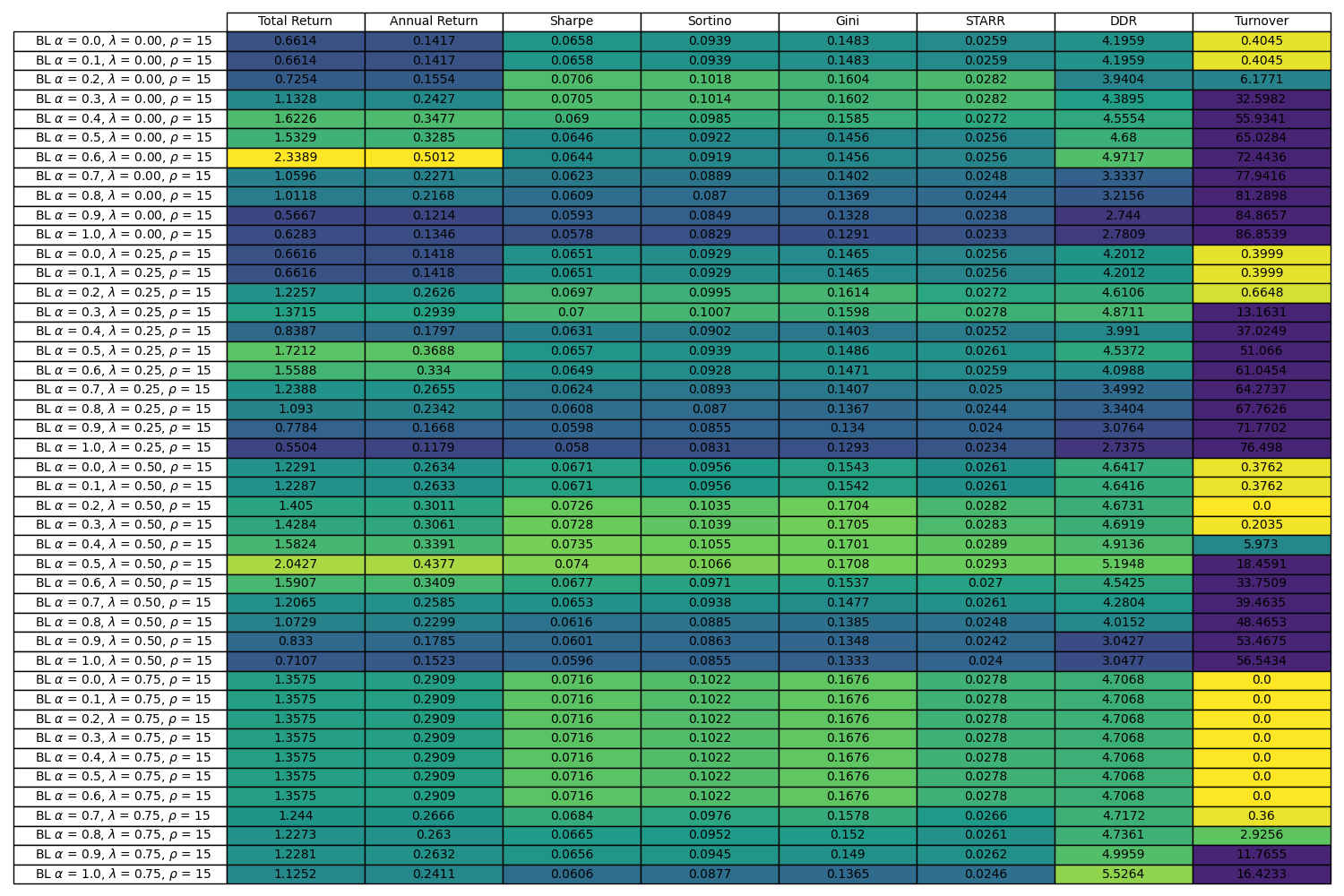}
    \caption{}
    \label{C95_BL15}
\end{figure}

\renewcommand{\thefigure}{2e}
\begin{figure}[b]
    \centering
    \includegraphics[width = 19.5cm, angle = 90]{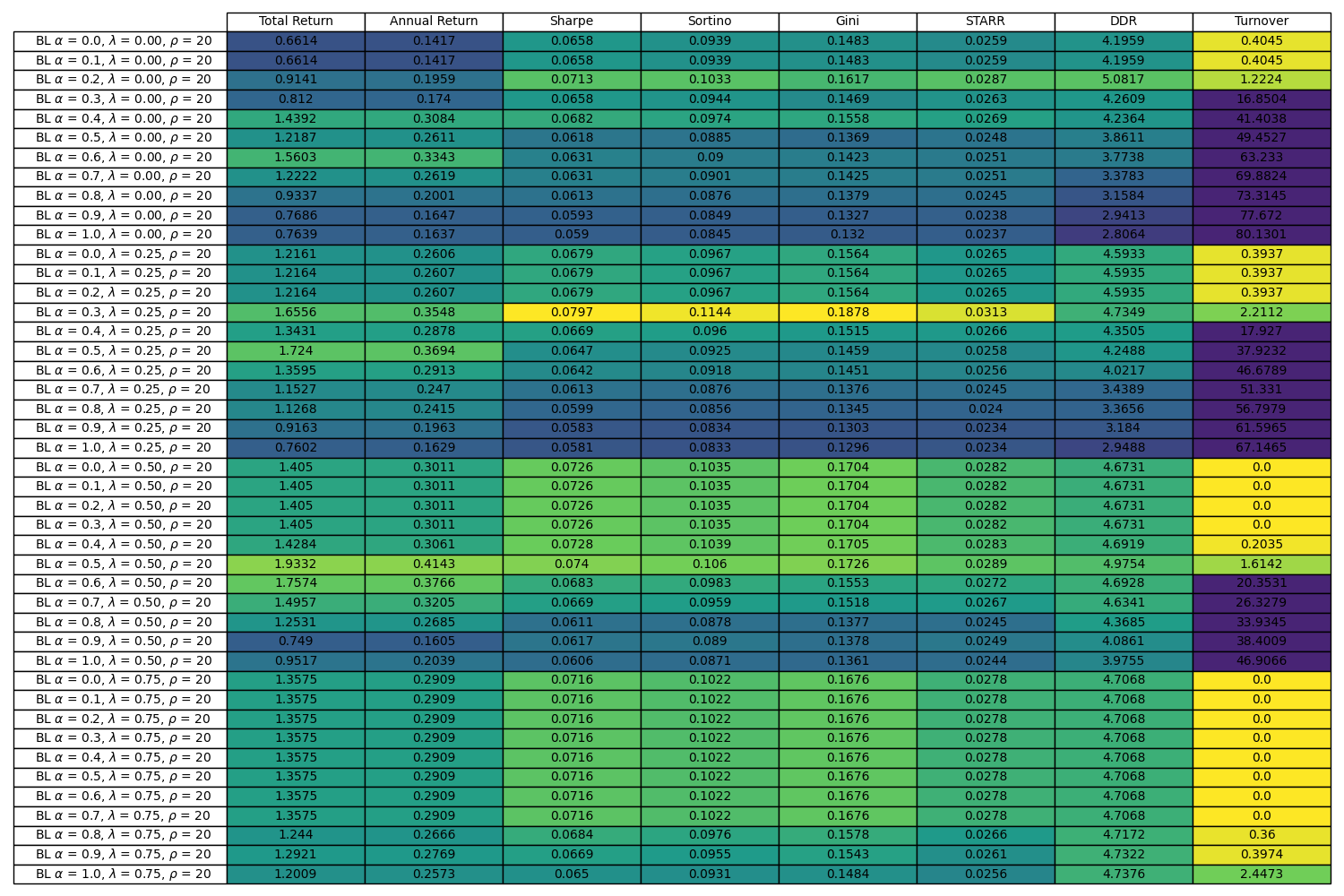}
    \caption{}
    \label{C95_BL20}
\end{figure}

\renewcommand{\thefigure}{2f}
\begin{figure}[b]
    \centering
    \includegraphics[width = 19.5cm, angle = 90]{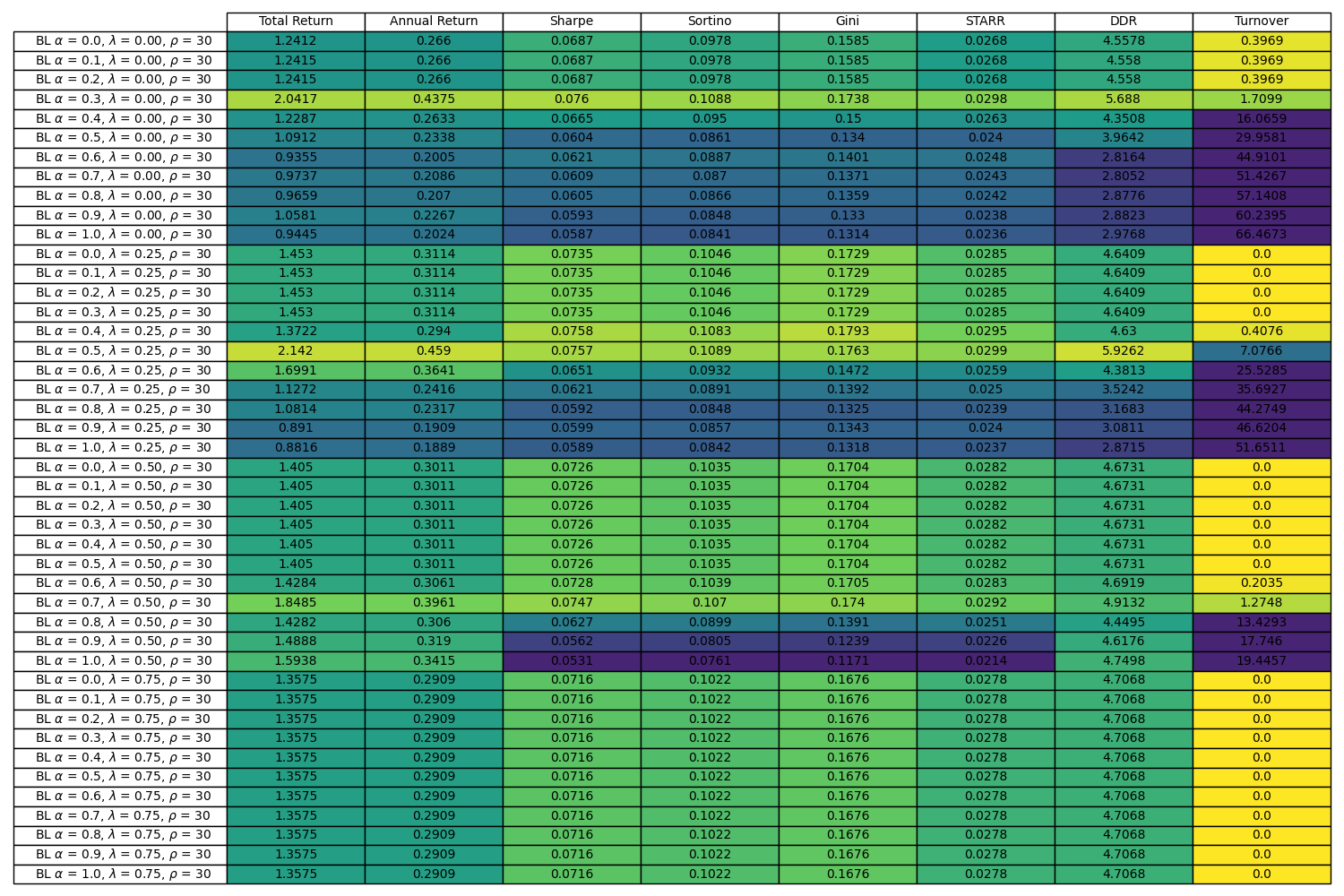}
    \caption{}
    \label{C95_BL30}
\end{figure}

\renewcommand{\thefigure}{2g}
\begin{figure}[b]
    \centering
    \includegraphics[width = 19.5cm, angle = 90]{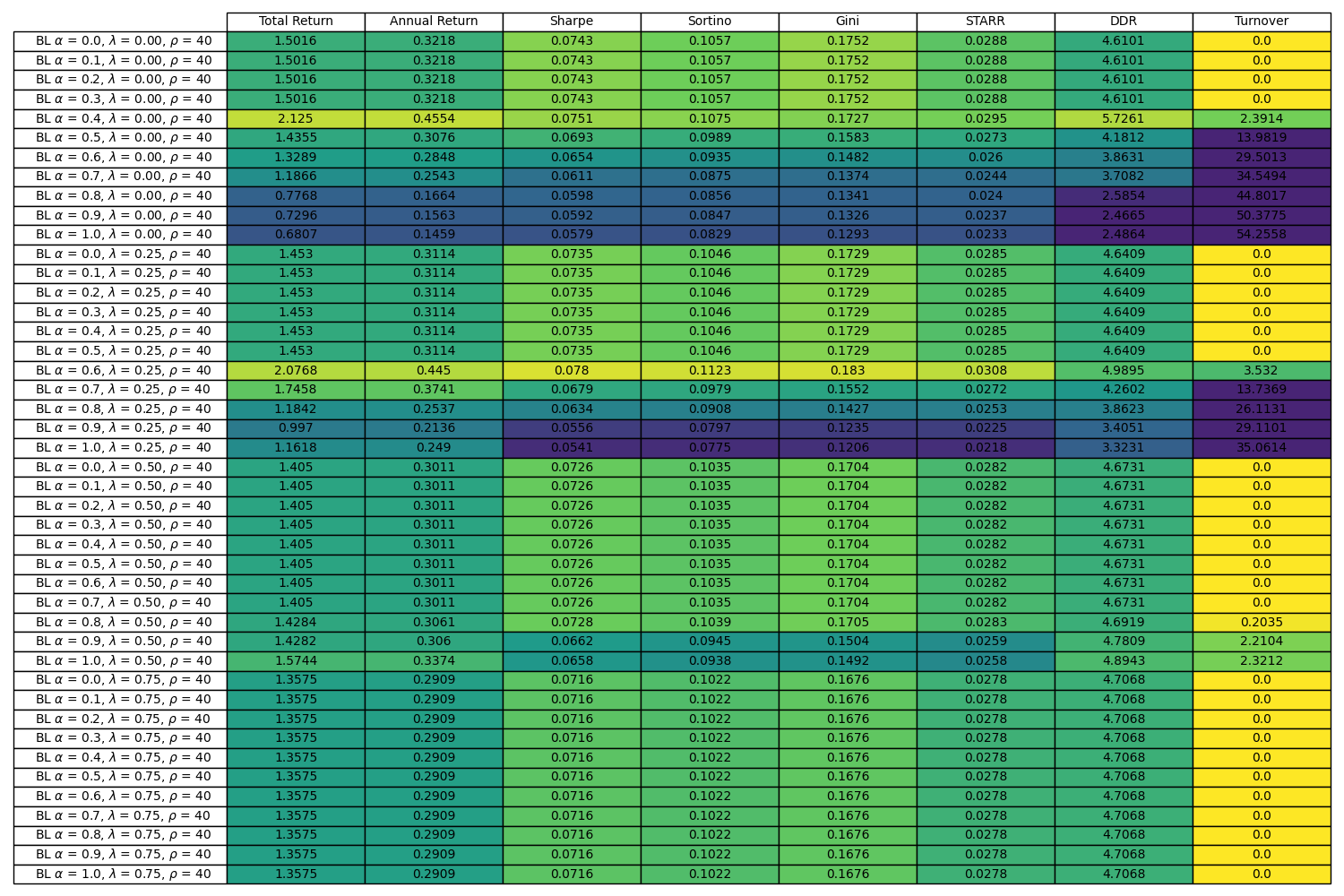}
    \caption{}
    \label{C95_BL40}
\end{figure}

\renewcommand{\thefigure}{3a}
\begin{figure}[b]
    \centering
    \caption{Tables 3A-3G display performance metrics of each portfolio strategy using CVaR99. Similar to the above tables linear gradients display the max and min of a column across the rows of all tables. The gradient for turnovers is in the same fashion as describe in Table 2A.}
    \includegraphics[width = 19.5cm, angle = 90]{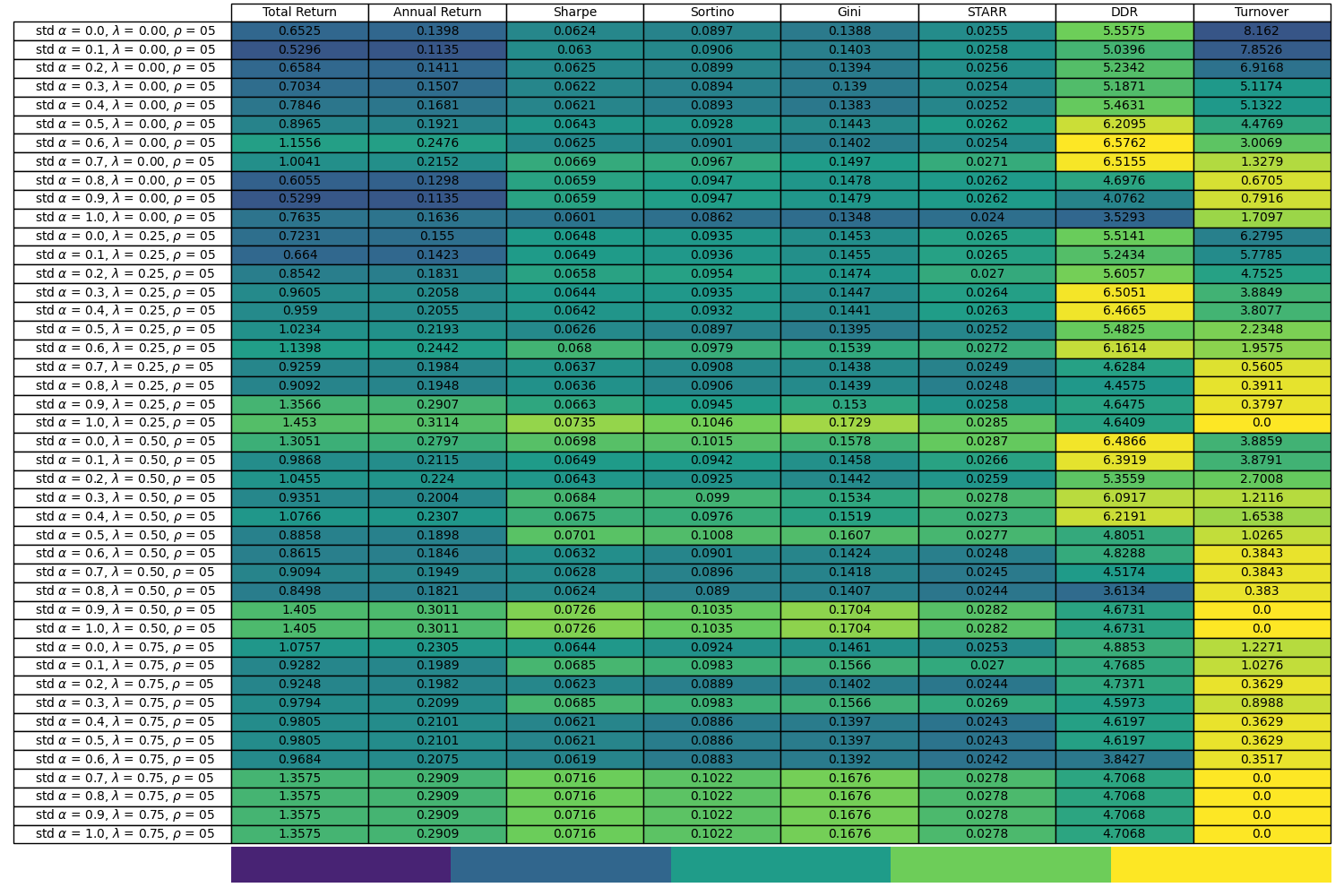}
    \caption{}
    \label{C99_std05}
\end{figure}

\renewcommand{\thefigure}{3b}
\begin{figure}[b]
    \centering
    \includegraphics[width = 19.5cm, angle = 90]{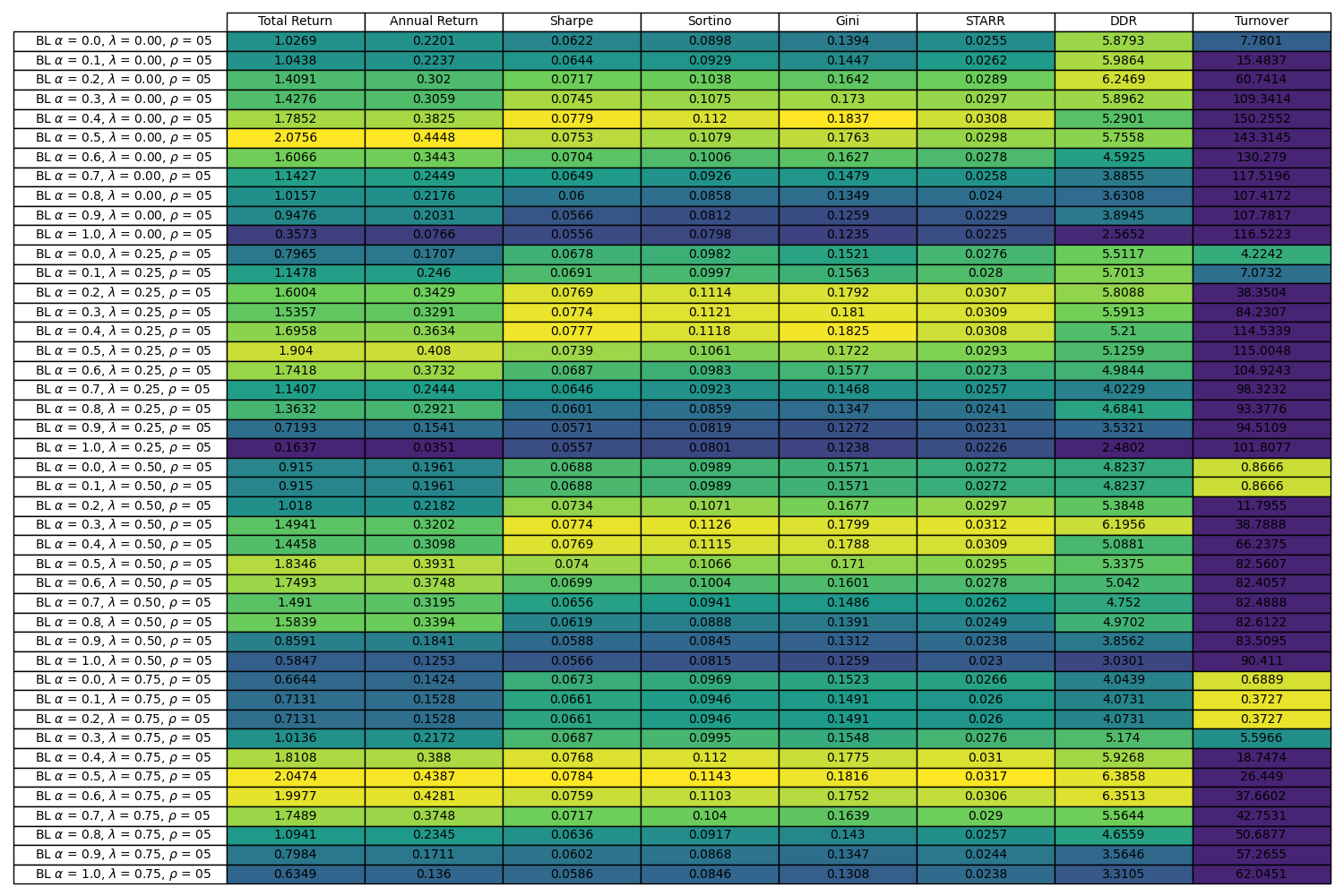}
    \caption{}
    \label{C99_BL05}
\end{figure}

\renewcommand{\thefigure}{3c}
\begin{figure}[b]
    \centering
    \includegraphics[width = 19.5cm, angle = 90]{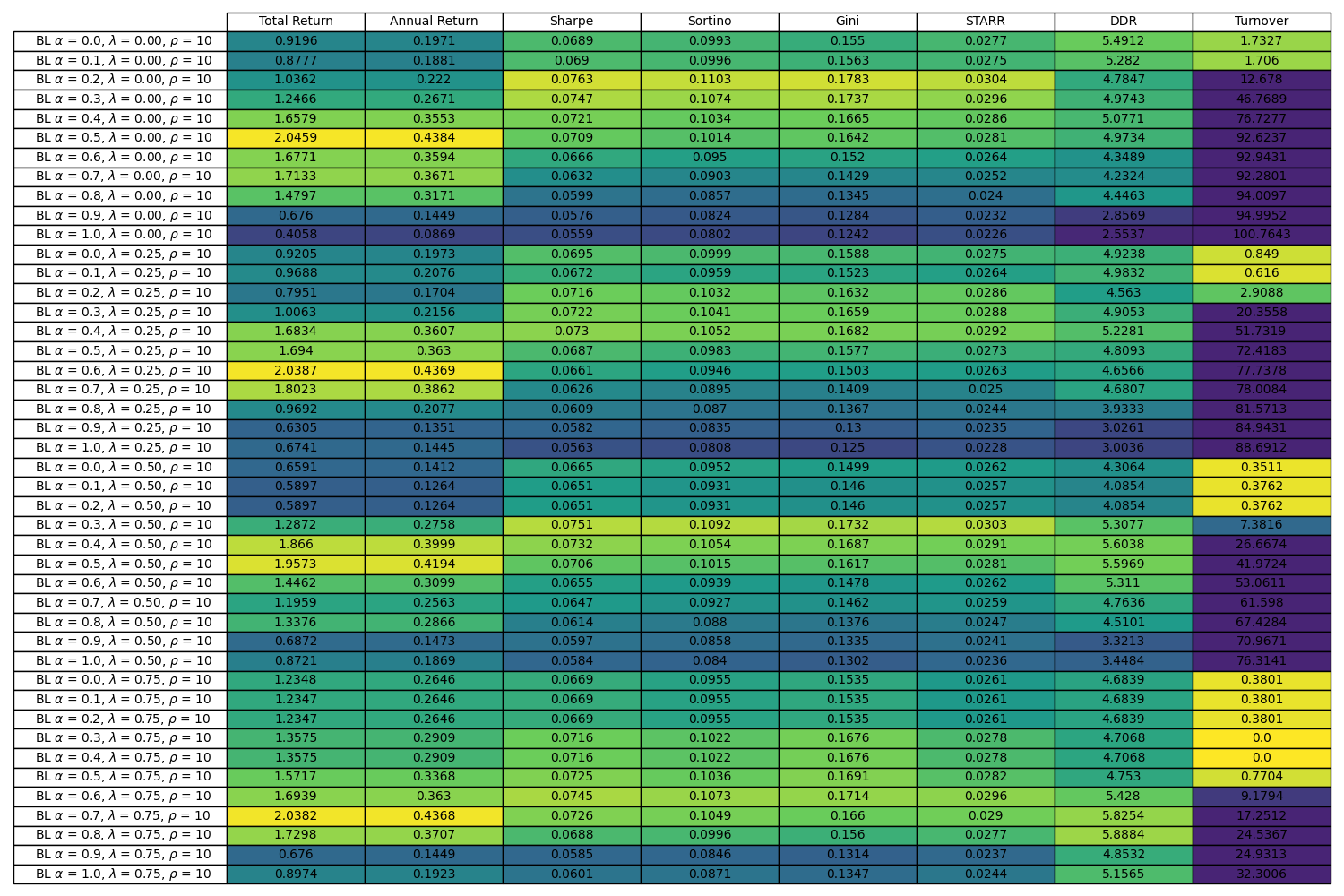}
    \caption{}
    \label{C99_BL10}
\end{figure}

\renewcommand{\thefigure}{3d}
\begin{figure}[b]
    \centering
    \includegraphics[width = 19.5cm, angle = 90]{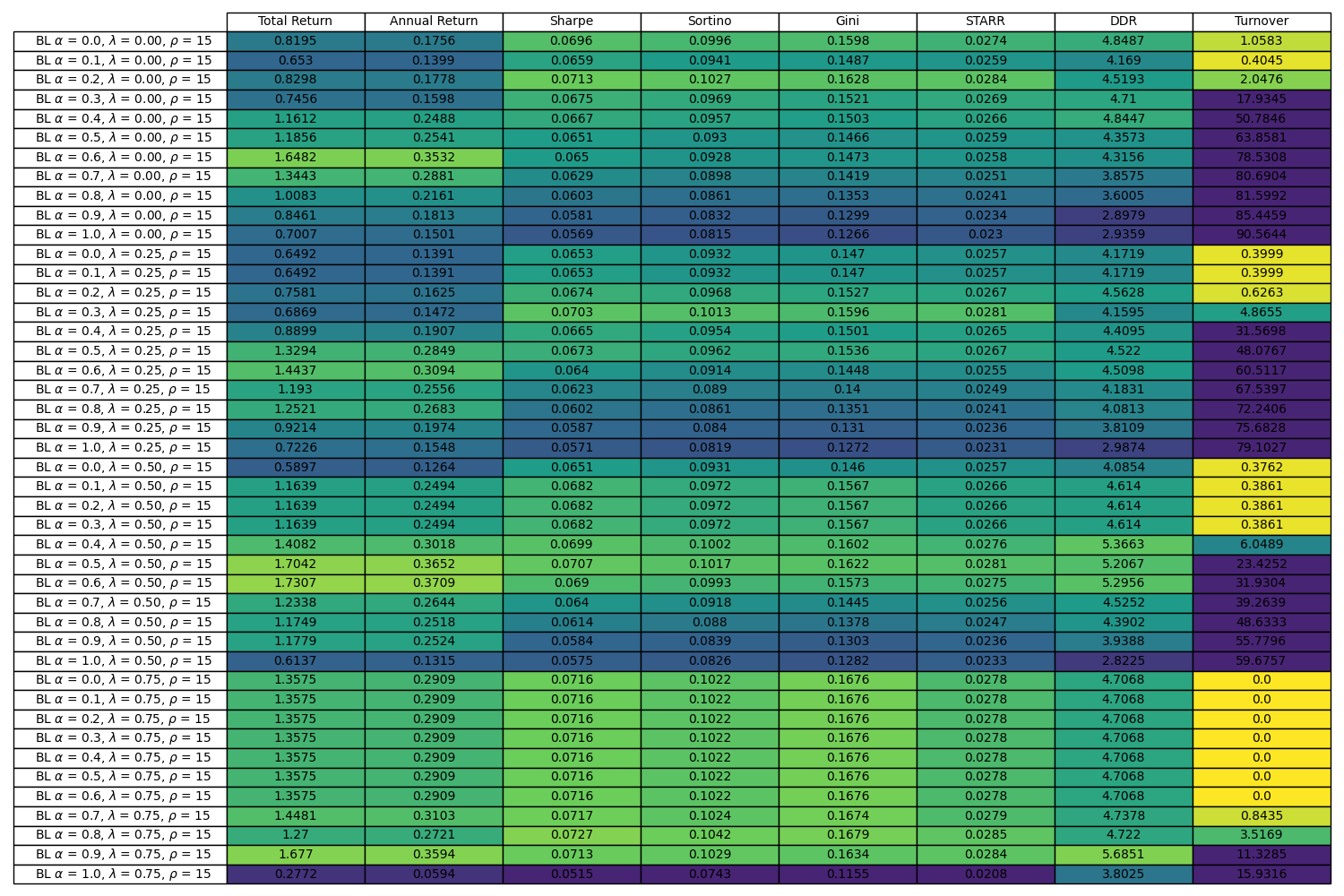}
    \caption{}
    \label{C99_BL15}
\end{figure}

\renewcommand{\thefigure}{3e}
\begin{figure}[b]
    \centering
    \includegraphics[width = 19.5cm, angle = 90]{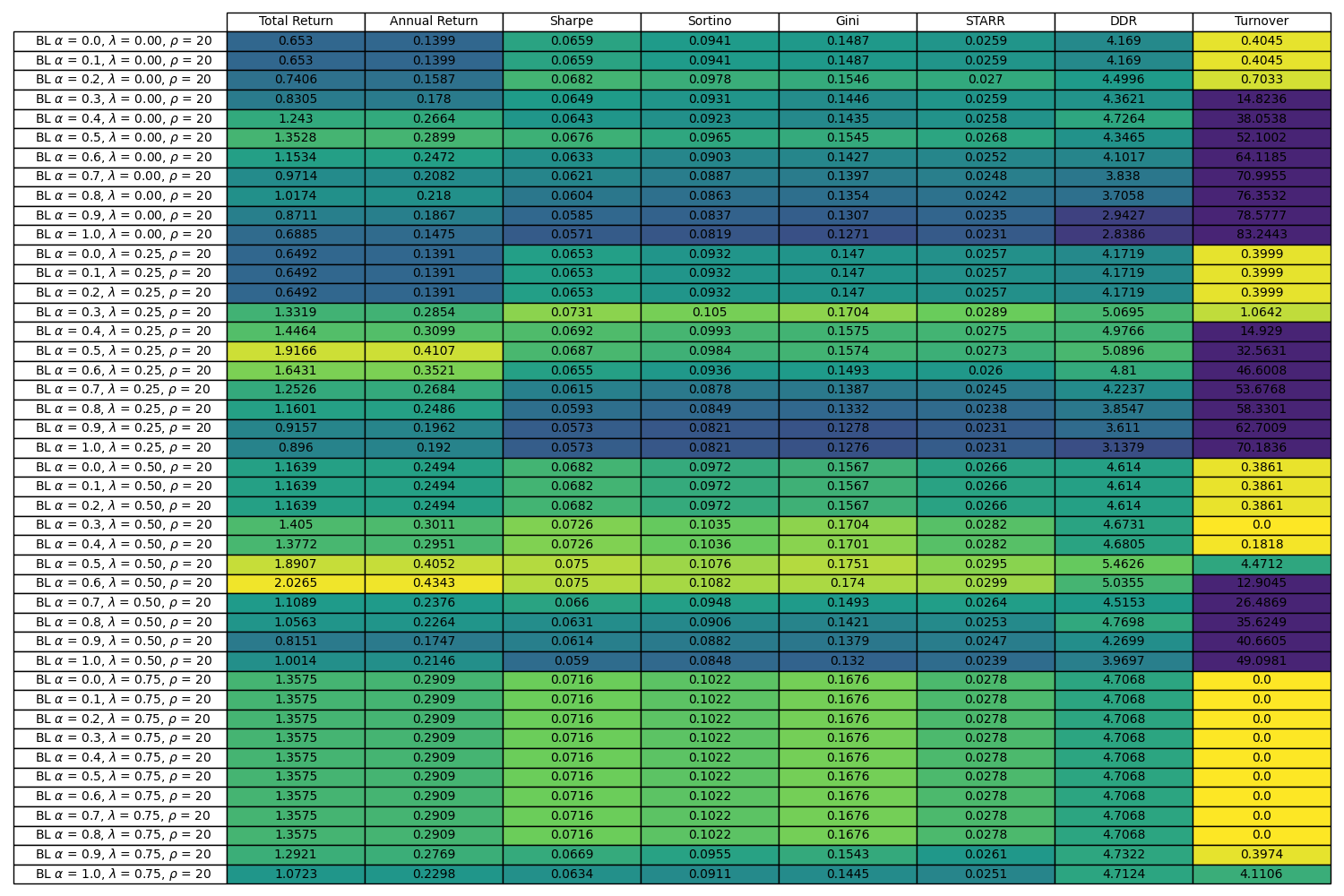}
    \caption{}
    \label{C99_BL20}
\end{figure}

\renewcommand{\thefigure}{3f}
\begin{landscape}
\begin{figure}[b]
    \centering
    \includegraphics[width = 19.5cm, angle = 90]{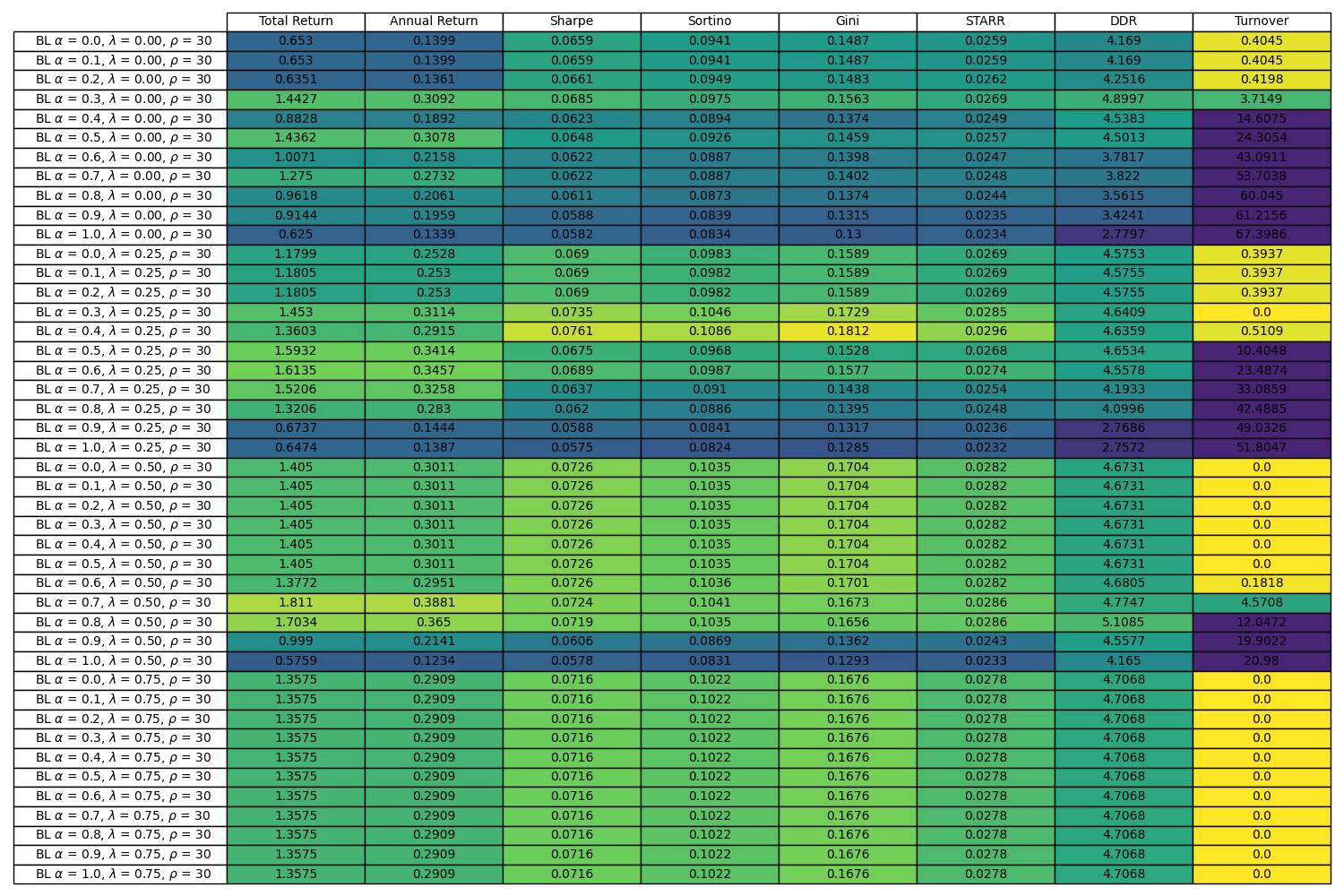}
    \caption{}
    \label{C99_BL30}
\end{figure}
\end{landscape}

\renewcommand{\thefigure}{3g}
\begin{landscape}
\begin{figure}[b]
    \centering
    \includegraphics[width = 19.5cm, angle = 90]{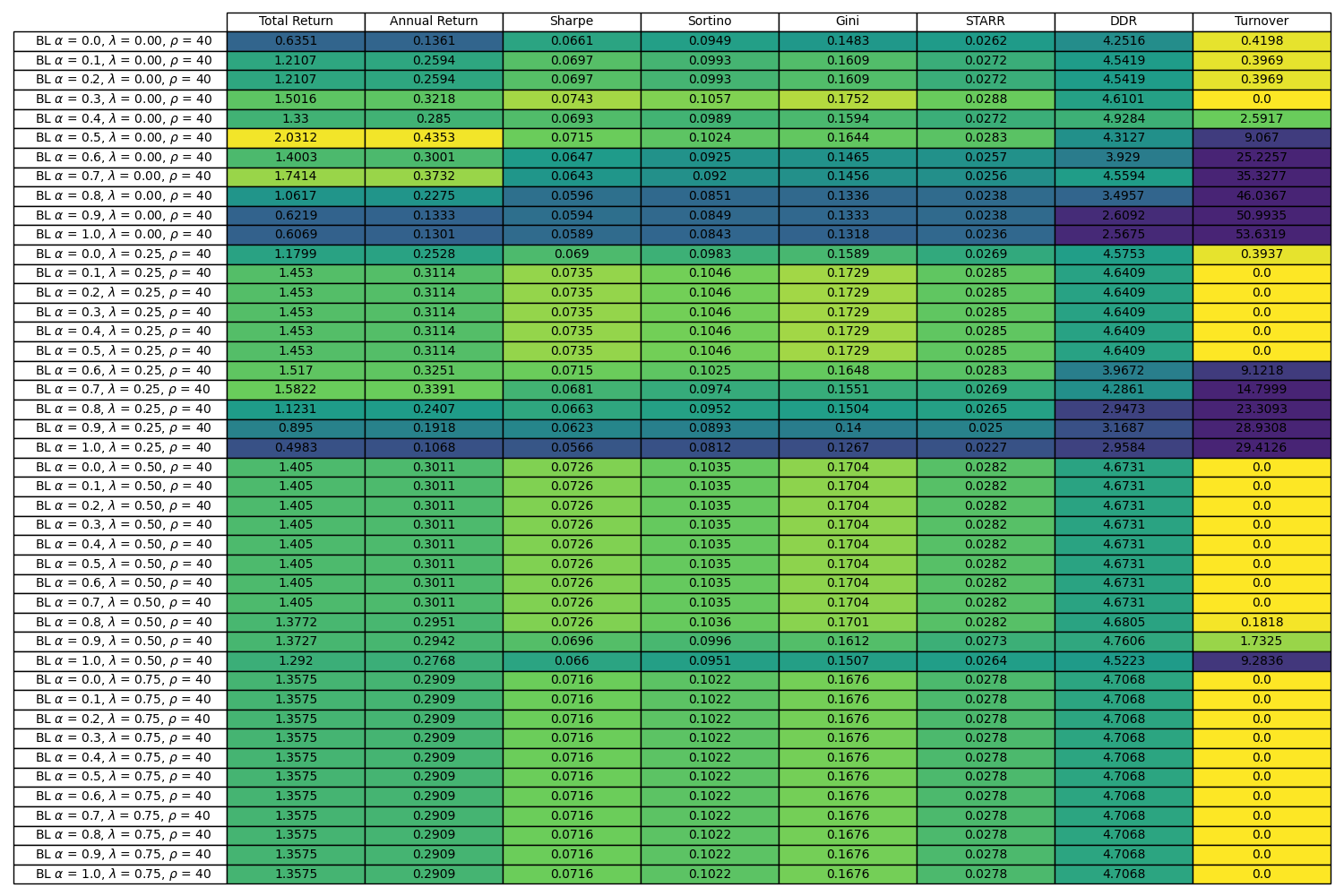}
    \caption{}
    \label{C99_BL40}
\end{figure}
\end{landscape}

\end{document}